\documentstyle[prl,preprint,aps,epsf]{revtex}
\newcommand{\beq}{\begin{equation}}
\newcommand{\eeq}{\end{equation}}
\newcommand{\beqs}{\begin{eqnarray}}
\newcommand{\eeqs}{\end{eqnarray}}

\newcommand{\nn}{\nonumber}

\begin{document}
\preprint{GTP-96-03}
\title
{Test of dilute gas approximation in quantum mechanical model}
\author
{D.K.Park$^a$,Soo-Young Lee$^b$, Jae-Rok Kahng$^b$, Sahng-Kyoon Yoo$^c$,  
C.H.Lee$^d$,\\ 
Chang Soo Park$^e$,
Eui-Soon Yim$^f$}
\address
{$^a$ Department of Physics, Kyungnam University, Masan 631-701, Korea \\
$^b$ Department of Physics, College of Science, Korea University,
Seoul 136-701, Korea \\
$^c$ Department of Physics, Seonam University, Namwon, Chunbuk 590-170, Korea \\
$^d$ D\&S Dept.,R\&D Center, Anam Industrial CO., LTD, Seoul 133-120, Korea \\ 
$^e$ Department of Physics, Dankook University, Cheonan  330-180, Korea \\
$^f$ Department of Physics, Semyung University, Chechon 390-230, Korea}
\date{\today}
\maketitle
\begin{abstract}
\indent 
The validity of dilute gas approximation is explored by making use of
the large-sized instanton in quantum mechanical model. It is shown that
the Euclidean probability amplitude derived through a dilute gas approximation
not only cannot explain the result of the linear combination of atomic orbitals 
approximation, but also does not exhibit a proper limiting case when the size
of instanton is very large.
\end{abstract}

\newpage
It is well known that instanton method is a useful tool for the quantitative
understanding of quantum tunneling[1, 2]. Although this method cannot 
inevitably avoid the Gaussian approximation, which is a common feature of 
path-integral formalism, in the course of evaluation of quantum fluctuation,
it is believed that the multi-instanton contribution compensates the 
Euclidean probability amplitude for the information loss resulted from 
this approximation to a large extent.
First successful result about this is reported by 
E. Gildener and A. Patrascioiu[3] in the
quantum tunneling problem of double well potential. After treating
the zero mode carefully, they calculated the multi-instanton contribution
by using a dilute gas approximation. Comparing their result with that
of linear combination of atomic orbitals(LCAO) approximation[4],  
they proved that their method provides a reasonable energy
splitting, $\Delta E$, due to tunneling effect. Recently it is shown that 
the repeat 
of the same procedure in a triple well potential case also gives a LCAO-type  
energy splitting[5].

Giving a very reasonable result in the quantum mechanical tunneling 
problem, the dilute instanton gas approximation gets some doubt in the 
scale invariant theory like the real dimensional QCD. This is because
scale invariance makes the instanton size not to be determined.
In the real dimensional QCD the large size of instanton gives rise 
to a troublesome infrared problem in the $\theta$-vacuum energy.
Although several different approaches[6, 7, 8] are suggested in order to 
escape from this difficulty, the result does not seem to be conclusive.

In this paper we will examine the validity of a dilute gas approximation
and explore the information loss due to this approximation by introducing
the large-sized instanton in the simple quantum mechanical model.
The large-sized instanton, which is essential for the test of the dilute
gas approximation, is easily produced by using a potential
\beq
V(\phi) = \alpha (\phi^2 + \gamma) (\phi - \beta)^2 (\phi + \beta)^2
\eeq
where $\gamma > 0$.  
An interesting feature of the potential (1) is that it can
be either double well or triple well depending on the values of $\gamma$
and $\beta$ as shown in Fig.1.
In the double well region ($ \gamma > \frac{\beta^2}{2} $), it is well
known that the LCAO approximation results in 
\beq
\lim_{T \rightarrow \infty} <\phi_f, T \mid \phi_i, -T>
\propto <\phi_f \mid R_0> <L_0 \mid \phi_i> e^{-2 E_0 T} \sinh(2 \Delta E T)
\eeq
where $E_0$ and $\Delta E$ are the unperturbed vacuum energy and half of energy
splitting, respectively, and $\mid R_0 >$ and $\mid L_0 >$ denote the 
normalized lowest energy eigenstates of the isolated wells.
For the small $\gamma$ case($ \gamma \ll \frac{\beta^2}{2} $), 
the central false vacuum holds off the transition from left true vacuum
to right one and makes the large-sized instanton as a result of this fact.
The smaller $\gamma$ is, the larger instanton we can get. 
In this case the LCAO approximation gives
\beqs
& &\lim_{T \rightarrow \infty} <\phi_f, T \mid \phi_i, -T>  \\  \nn
&=& <\phi_f \mid R_0 > <L_0 \mid \phi_i>
  \left[ \frac{e^{-2 E'_0 T}}{2 + a_+^2} + 
         \frac{e^{-2 E'_2 T}}{2 + a_-^2} - 
         \frac{1}{2} e^{-2 E'_1 T}
  \right]
\eeqs
where $a_{\pm}$ and $E'_n (n = 0, 1, 2)$ are given in Ref.[5].
So it is very easy to show that 
$\lim_{T \rightarrow \infty} <\phi_f, T \mid \phi_i, -T>$ cannot contain
the hyperbolic-sine term in this case.

Now let us calculate the same quantity by following the 
usual instantonic procedure.
By integrating the Euclidean equation of motion
\beq
\frac{1}{2} (\frac{d \phi}{d \tau})^2 = \alpha (\phi^2 + \gamma)
                                        (\phi - \beta)^2
                                        (\phi + \beta)^2
\eeq
the instanton solution and corresponding classical Euclidean action for 
large $T$ are straightforwardly derived after a tedious calculation:
\beqs
\phi_{cl}&=& \epsilon(\tau - \tau_0) \beta
            \sqrt{
                   \frac{\cosh[2 \sqrt{2 \alpha ( \beta^2 + \gamma) } \beta
                              (\tau - \tau_0) ] - 1}
                        {\cosh[2 \sqrt{2 \alpha ( \beta^2 + \gamma) } \beta
                              (\tau - \tau_0)] + \frac{2 \beta^2 + \gamma}
                                                      {\gamma}  }
                  }        \\    \nn
S_{cl}&=& \sqrt{\alpha} \beta^3 \frac{(\beta^2 + \gamma)^{5/2}}{\gamma^2}
         \frac{\Gamma(\frac{11}{4})}{\Gamma(\frac{5}{2}) \Gamma(\frac{1}{4})}
         F(3, \frac{5}{2}; \frac{11}{4}; -\frac{\beta^2}{\gamma})
\eeqs
where $\epsilon(x)$ and $F(a, b; c; z)$ are usual alternating
and hypergeometric functions.

By defining the change of variable
\beq
\phi = \phi_{cl} + \eta
\eeq
one instanton contribution to the Euclidean probability amplitude 
becomes
\beq
<\phi_f, T \mid \phi_i, -T > = e^{-S_{cl}} I_1(T)
\eeq
where
\beq
I_1(T) = \int_{(-T, 0)}^{(T, 0)} D \eta e^{-\int d \tau \eta \hat{M} \eta}
\eeq
and 
\beq
\hat{M} = - \frac{1}{2} \frac{d^2}{d \tau^2} + \alpha
        [15 \phi_{cl}^4 - 6 (2 \beta^2 - \gamma) \phi_{cl}^2
         + \beta^2 ( \beta^2 - 2 \gamma) ].
\eeq
It is straightforward to prove that $\dot{\phi_{cl}} \equiv \frac{d \phi_{cl}}
{d \tau}$ is zero mode of $\hat{M}$. The repeat of the procedure given in 
Ref.[3] enables one to get the one instanton contribution of the 
Euclidean amplitude as
\beq
<\phi_f, T \mid \phi_i, -T>_{(1)} = 
e^{-S_{cl}} \frac{16 T}{\pi}
            \frac{\beta^3 (\beta^2 + \gamma)^2 2 \alpha}{\gamma}
            e^{-2 \sqrt{2 \alpha (\beta^2 + \gamma)} \beta T}
\eeq
where the subscript $(1)$ means the one instanton contribution.

The multi-instanton contribution to the probability amplitude can be obtained 
by summing the possible configurations and using a dilute gas approximation:
\beqs
& &<\phi_f, T \mid \phi_i, -T>    \\  \nn
&=&\sqrt{\frac{\omega}{\pi}} e^{-\omega T}
   \sum_{n = 0}^{\infty}\frac{1}{(2 n + 1)!} 
                        \left[ e^{-S_{cl}} \frac{16 T}{\sqrt{\pi \omega}}
                                \frac{\beta^3 (\beta^2 + \gamma)^2 2 \alpha}
                                     {\gamma}
                         \right]^{2n+1}   \\  \nn 
&=&\sqrt{ \frac{ 2 \sqrt{2 \alpha (\beta^2 + \gamma)} \beta}{\pi} }
 e^{-2 \sqrt{2 \alpha (\beta^2 + \gamma)} \beta T}
 \sinh \left[ e^{-S_{cl}} \frac{16 T}{\sqrt{2 \pi}}
             \frac{(2 \alpha)^{3/4} \beta^{5/2} (\beta^2 + \gamma)^{7/4}}
                  {\gamma}
      \right]
\eeqs
where $\omega = 2 \sqrt{2 \alpha (\beta^2 + \gamma)} \beta$.
In Eq.(11) the most dominant instanton density in the summation is
\beq
\frac{n}{T} \sim e^{-\frac{S_{cl}}{\hbar}}
                 \frac{16}{\sqrt{\pi \omega}}
                 \frac{\beta^3 (\beta^2 + \gamma)^2 2 \alpha}{\gamma}
\eeq
where the Plank constant $\hbar$ in the exponent is explicitly expressed for the
following remark.

Since the classical Euclidean action $S_{cl}$ is finite in the 
double well region, the
dominant instanton density becomes very small, which guarantees the 
justification
of the dilute gas approximation.
By comparing Eq.(11) with (2) one can read
\beqs
E_0&=& \sqrt{2 \alpha ( \beta^2 + \gamma)} \beta      \\  \nn
\Delta E&=& \frac{8}{\sqrt{2 \pi}} 
           \frac{(2 \alpha)^{3/4} \beta^{5/2} (\beta^2 + \gamma)^{7/4}}
                {\gamma}
            e^{-S_{cl}}
\eeqs
in this region.
If $\gamma$ increases, $\Delta E \propto \gamma^{\frac{3}{4}} 
e^{-c_1 \gamma^{1/2}}$, where $c_1$ is some constant, goes to zero
which is physically reasonable.  

However, situation is quite different in the small $\gamma$ region.
In this region one can easily show
\beq
S_{cl} = c_2  \gamma^{\frac{1}{2}},
\eeq
where $c_2$ is a $\gamma$-independent constant, by invoking the asymptotic
formula of hypergeometric function. If $\gamma$ is less than or order of
$(\hbar / c_2)^2$, Eq.(12) casts a doubt on the validity of the dilute gas 
approximation
in this region. As expected, the probability amplitude derived through
dilute gas approximation  
\beq
<\phi_f, T \mid \phi_i, -T> \propto
\sinh \left[
            e^{-S_{cl}} \frac{16 T}{\sqrt{2 \pi}}
                \frac{(2 \alpha)^{3/4} \beta^6}{\gamma}
     \right].
\eeq
cannot explain the LCAO result (3). The worse fact in this region is that 
$\gamma = 0$ limit of Eq.(15) is not well-defined although the potential
is rarely changed from $\alpha \phi^2 (\phi - \beta)^2 (\phi + \beta)^2$.
This means that multi-instanton contribution calculated through the
dilute gas approximation loses too much information when the size of 
instanton is very large. Although one can see the breakdown of dilute
gas approximation in the strong coupling region of the double well
case[9], the origin of the breakdown is quite different in this case.
The origin of this phenomena in this case seems to be the sensitivity
and large size of classical solution.

In this stage the following natural and fundamental question arises:
Is there any other approximation method which replaces the dilute 
gas approximation
and is able to explain LCAO result regardless of the instanton size?
Although the model presented in this paper is not scale invariant,
the investigation of the answer of the above question in this simple 
quantum mechanical model may shed light
on the complicated QCD problem.

\vspace{.5cm}

 ACKNOWLEDGEMENT 

Some of the authors would like to approciate the partial support by
Nondirected Research Fund, Korea Research Foundation,
'93 and '95(E.S.Y and S.K.Y), and by Korea Science and Engeneering
Foundation (961-0201-005-1)

\begin{figure}
\caption{potential (1) at double well and triple well regions}
\end{figure}

\newpage
\epsfysize=20cm \epsfbox{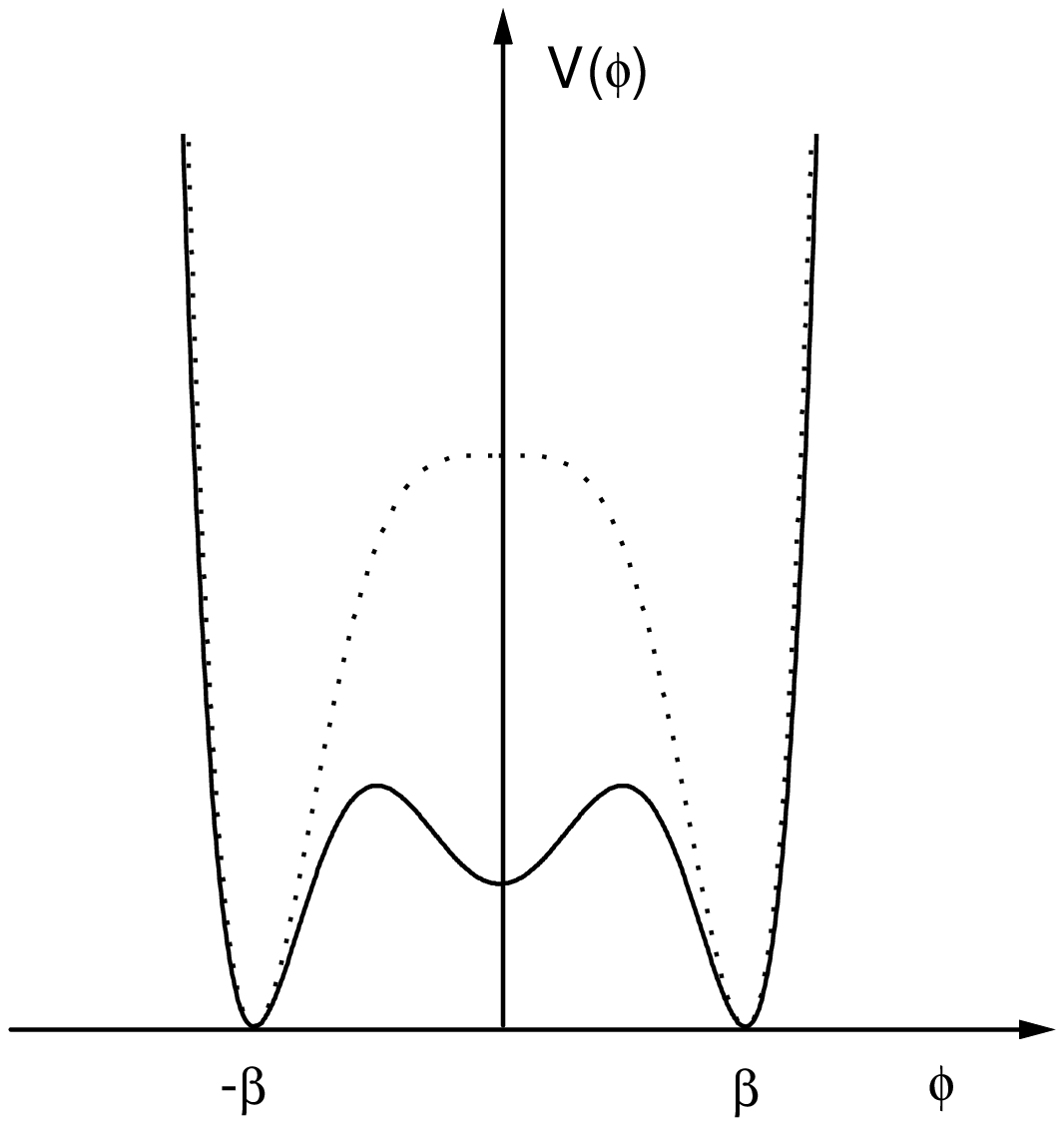}


\normalsize
\begin{thebibliography}{99}
\bibitem{1} R.Rajaraman, {\it Solitons and Instantons}, (North-Holland, the Netherlands, 1987).
\bibitem{2} S.Coleman, "The uses of Instantons",
in {\it The why's of subnuclear physics},
ed. A.Zichichi (Plenum, New York, 1979).
\bibitem{3} E.Gildener and A.Patrascioiu, Phys. Rev. D{\bf 16}, 423 (1977).
\bibitem{4} R.L.Liboff, 
{\it Introductory Quantum Mechanics}, (Addison-Wesley Publishing Company, Inc. 1992).
\bibitem{5} Soo-Young Lee, Jae-Rok Kahng, Sahng-Kyoon Yoo, D.K.Park, C.H.Lee, Chang Soo Park and Eui-Soon Yim, quant-ph/9608015.
\bibitem{6} C.G.Callan, R.F.Dashen, D.J.Gross, Phys. Lett. B{\bf 66}, 334 (1977);Phys. Rev. D{\bf 17}, 2717 (1978).
\bibitem{7} V.A.Fateev, I.V.Frolov and A.S.Schwarz, Nucl. Phys. B{\bf 154}, 1 (1979).
\bibitem{8} B.Berg and M.Luscher, Comm. Math. Phys. {\bf 69}, 57 (1979); Nucl. Phys. B{\bf 160}, 281 (1979).
\bibitem{9} J.F.Willemsen, Phys. Rev. D{\bf 20}, 3292 (1979).




\end{thebibliography}
\end{document}